# On the room temperature multiferroic BiFeO$_3$: Magnetic, dielectric and thermal properties

Jun Lu,[1,2] A. Günther,[1] F. Schrettle,[1] F. Mayr,[1] S. Krohns,[1] P. Lunkenheimer,[1,a] A. Pimenov,[3] V. D. Travkin,[4] A. A. Mukhin,[4] A. Loidl[1]

[1] Experimental Physics V, Center for Electronic Correlations and Magnetism, University of Augsburg, 86159 Augsburg, Germany
[2] School of Materials Science and Engineering, University of Science and Technology, Beijing 100083, China
[3] Experimentelle Physik IV, Universität Würzburg, 97074 Würzburg, Germany
[4] General Physics Institute of the Russian Academy of Sciences, 119991 Moscow, Russia

**Abstract.** Magnetic dc susceptibility between 1.5 and 800 K, ac susceptibility and magnetization, thermodynamic properties, temperature dependence of radio and audio-wave dielectric constants and conductivity, contact-free dielectric constants at mm-wavelengths, as well as ferroelectric polarization are reported for single crystalline BiFeO$_3$. A well developed anomaly in the magnetic susceptibility signals the onset of antiferromagnetic order close to 635 K. Beside this anomaly no further indications of phase or glass transitions are indicated in the magnetic dc and ac susceptibilities down to the lowest temperatures. The heat capacity has been measured from 2 K up to room temperature and significant contributions from magnon excitations have been detected. From the low-temperature heat capacity an anisotropy gap of the magnon modes of the order of 6 meV has been determined. The dielectric constants measured in standard two-point configuration are dominated by Maxwell-Wagner like effects for temperatures T > 300 K and frequencies below 1 MHz. At lower temperatures the temperature dependence of the dielectric constant and loss reveals no anomalies outside the experimental errors, indicating neither phase transitions nor strong spin phonon coupling. The temperature dependence of the dielectric constant was measured contact free at microwave frequencies. At room temperature the dielectric constant has an intrinsic value of 53. The loss is substantial and strongly frequency dependent indicating the predominance of hopping conductivity. Finally, in small thin samples we were able to measure the ferroelectric polarization between 10 and 200 K. The saturation polarization is of the order of 40 µC/cm$^2$, comparable to reports in literature.

## 1 Introduction

After extensive work in the late sixties [1], in recent years multiferroics gained considerable attention driven by the progress that has been achieved in sample preparation, measuring techniques and theoretical concepts [2,3]. Multiferroics exhibiting simultaneous polar and magnetic order promise important applications, as in cases of strong magneto-electric coupling both the magnetisation and the polarisation can be manipulated by electric as well as by magnetic fields. BiFeO$_3$ (BFO) is a rhombohedrally distorted perovskite with space group R3c and belongs to the rare class of materials with long range magnetic and long range ferroelectric (FE) order already at room temperature [4]. After early work which mainly was dedicated to clarify structural, polar and magnetic properties of BFO, a revival has been triggered by the general focus on multiferroics in the materials-based community and by the observation of weak ferromagnetism coexisting with ferroelectricity in thin films [4]. However, BFO can hardly be synthesized in pure and stoichiometric form and many experiments suffer from impurity phases, magnetic defects or deviations from the ideal oxygen stoichiometry mimicking weak ferromagnetism and a number of further magnetic or structural phase transitions at low temperatures. In addition, this accidental doping leads to significant conductivity contributions already at room temperature, making the detection of hysteresis loops difficult and influencing measurements of dielectric constants with significant Maxwell-Wagner type contributions.

Despite these drawbacks and experimental complications the basic structural and magnetic properties of bulk BFO are nowadays well established: BFO undergoes polar order close to 1100 K [5] and becomes antiferromagnetic (AFM) at about 645 K [6,7]. Probably related to sample quality there is an enormous scatter of reported values of the AFM ordering temperature ranging between 595 K and 650 K [8]. At room temperature BFO crystallizes in a rhombohedrally distorted perovskite structure [9], which can be indexed in a rhombohedral cell with two formula units or in a hexagonal unit cell with six formula units. R3c symmetry allows for weak ferromagnetism (FM) and for spontaneous polarization along [111] as Bi, Fe and O are displaced along the threefold axis. It is clear that the stereochemically active Bi lone pair plays an essential role in establishing high-temperature ferroelectricty. The rhombohedral cell is derived from the perovskite structure by a counter rotation of the oxygen octahedra around the trigonal [111] axis. Detailed structural investigations of BFO in a broad temperature range from room temperature almost to the melting point have been performed by Bucci *et al.* [10] and by Palewicz *et al.* [11]. The essential outcome of the latter investigation was the observation that the shifts of Bi$^{3+}$ and Fe$^{3+}$ with respect to the ideal perovskite structure, which mainly determine the electric polarization of BFO, do not change significantly with temperature. Based on their data one expects a change of the polarization of the order of 6% between room temperature and the onset of ferroelectric order only [11]. Antiferromagnetic order with a G-type spin configuration below 650 K has been established by Kiselev et al. [7]. Using polarization analysis the magnetic moment has been determined as gS = 3.68 $\mu_B$ [12]. Later on, using high-resolution time-of-flight neutron scattering techniques it has been concluded that the collinear G-type structure is modified by a long-range modulation and that the ground state of BFO can be represented as cycloidal spiral with a period of approximately 62 nm [13]. Further neutron scattering diffraction demonstrated that the modulated magnetic ground state of BFO does not change much on cooling from the magnetic phase transition at 640 K down to 4 K [14]. High-pulsed magnetic field investigations revealed a suppression of the cycloidal spin structure at ~ 20 T which was accompanied by a jump of the electric polarization and an induced weak ferromagnetic moment [15].

[a] e-mail: peter.lunkenheimer@physik.uni-augsburg.de

Recently detailed investigations of the bulk magnetic properties of high quality single crystals of BFO have been reported [16]. Below 150 K the authors found a constant and almost temperature independent magnetization of the order of 60 emu/mol, as measured at 1 T, without traces of weak ferromagnetism. The room temperature magnetization as measured in external fields below 5 T was absolutely linear with field, typical for an antiferromagnetic spin arrangement [16]. These experimental observations outlined above contradict numerous reports of further structural or magnetic phase transitions in BFO below the magnetic ordering temperature at 640K [17,18,19,20,21,22,23]. It has been speculated that these low-temperature anomalies correspond to spin-reorientation or spin glass transitions. However, many of these anomalies do not coincide in temperature but rather point towards impurity phases or magnetic defects.

A FE hysteresis in BFO was measured for the first time by Teague et al. [24] at room temperature. These authors found a FE polarization of 3.5 $\mu C/cm^2$ with electric fields of approximately 50 kV/cm. FE hysteresis loops on thin films of pure and doped BFO have also been reported by Naganuma et al. [25] and on bulk samples by Lebeugle et al. [16] The latter found a polarization of almost 30 $\mu C/cm^2$ from which they deduced a full saturation polarization of 60 $\mu C/cm^2$ along the hexagonal [001] direction of BFO, a value which so far has only been detected in thin films.

Krainik et al. [26] measured the temperature dependence of the dielectric constant from room temperature up to 1150 K at microwave frequencies. They found a linear increase of the dielectric constant from a room temperature value $\varepsilon$ = 45 to approximately 150 at 1150 K. The dielectric loss remained almost constant with a loss tangent of 0.09. From these measurements the authors deduced a FE transition temperature of 1120 K. Having a closer look at the experimental results, this temperature seems to be rather ill-defined and the anomaly at high temperature could equally well signal the decomposition of the sample. It is rather well established that at high temperatures, but well below the melting point, BFO decomposes into $Bi_2Fe_4O_9$ and $Fe_2O_3$. Since then a number of dielectric measurements has been performed on ceramics [6,21,27,28,29,30,31,32,33] as well as on thin films [34,35,36]. The intrinsic dielectric constants as measured at either relatively high frequencies or low temperatures scatter between 50 to 300 and most of these experiments reveal a rather strong unusual frequency dependence and probably suffer from Maxwell-Wagner like contributions to the dielectric constants, mainly resulting from non-negligible conductivities in addition to external (contacts) or internal (grain boundaries, domain walls, inhomogeneities) boundaries. These external or internal barrier layers lead to a dramatic enhancement of the dielectric constants as often found in a variety of oxide ceramics [37]. In recent dielectric measurements [21] a number of anomalies close to 50, 140 and 200 K have been interpreted as anomalies due to structural or magnetic (spin glass and spin reorientation) phase transitions. We want to point out that a number of the above mentioned problems and open questions were addressed in a recent review article on BFO by Catalan and Scott [23].

In this communication we document detailed experiments on the magnetic, dielectric and thermodynamic properties of BFO. Especially, magnetic measurements on a high quality single crystal reveal no evidence of low-temperature (T < $T_N$) phase or glass transitions in BFO. Heat-capacity measurements indicate the absence of any phase transitions below room temperature and, in addition, allow determining the anisotropy gap of magnon excitations. Finally, we determine the exact value of the dielectric constants between 70 K and 300 K by THz spectroscopy and additionally provide broadband dielectric data between 1.5 K and 600 K. In these measurements we found a rather unusual frequency dependence of the ac conductivity with different power laws at low and high frequencies.

## 2 Experimental details and sample characterization

Pure polycrystalline $BiFeO_3$ samples for temperature dependent x-ray diffraction measurements were synthesized by solid state reaction plus leaching. In a first step the 1:1 molar mixture of 99.999% $Bi_2O_3$ and 99.99% $Fe_2O_3$ was filled into alumina crucibles and was heated at 800 ºC for 48 hours. The reaction product was leached in a large volume of 10 vol.% $HNO_3$ three times and then washed in distilled water five times. For structural characterization the material was dried and ground to fine powder. Small crystals for various dielectric and magnetic measurements were grown using a flux method with alumina crucibles. Mixed powders with 3.5:1 molar ratio of 99.999% $Bi_2O_3$ and 99.99% $Fe_2O_3$ were ground in a mortar for two hours. The homogeneous mixtures were enclosed in alumina crucibles and then put into a furnace with the following time-temperature profile: 200 ºC/h heating up to 850 ºC followed by four hours soaking, 0.5 ºC/h cooling down to 750 ºC and finally furnace cooling. The crystals were then obtained via washing the flux product in 10 vol.% $HNO_3$ solution at around 90 ºC. After leaching with distilled water and drying, brown flat crystals were chosen and their dimensions were measured under optical microscopy revealing typical thicknesses of 10 - 40 $\mu m$ and typical areas of 0.1 - 0.4 $mm^2$. Before dielectric measurements all crystals were checked by SQUID measurements for possible impurity phases. Finally, we received a large single crystal (5 × 4 × 0.4 mm) from V.A. Murashov, which has been grown by flux methods as described elsewhere [38].

Ac and dc magnetic susceptibility and magnetization have been studied utilizing a commercial Superconducting Quantum Interference Device magnetometer (Quantum Design MPMS-5) with external magnetic fields up to 5 T. To check for possible impurity phases or for defect spins, we characterized all samples which have been used in this work by dc magnetization measurements. The heat capacity was measured in a Quantum Design Physical Properties Measurements System for temperatures between 1.8 K and 300 K. Here only the large single crystal has been investigated. The dielectric properties of the large and a number of small single crystals were determined using a frequency response analyzer (Novocontrol) at frequencies between 1 Hz and 1.5 MHz [39]. FE hysteresis loops were measured using a Ferroelectric Analyzer (TF2000, aixACCT) and a home built high-precision set up based on a Sawyer-Tower circuit. All dielectric measurements have been performed along the pseudocubic c-axis of BFO. Problems in measuring FE hystereses in leaky oxide samples are described by Scott [40] and Loidl *et al*. [41]. For these measurements silver paint or sputtered gold contacts were applied onto both sides of individual crystals. Finally, the dielectric constants of the large single crystal have also been measured using a millimetre-wave spectrometer [42] utilized in Mach-Zehnder geometry to determine real and imaginary part of the dielectric constant as independent quantities and, in addition and most importantly, in a contact-free configuration.

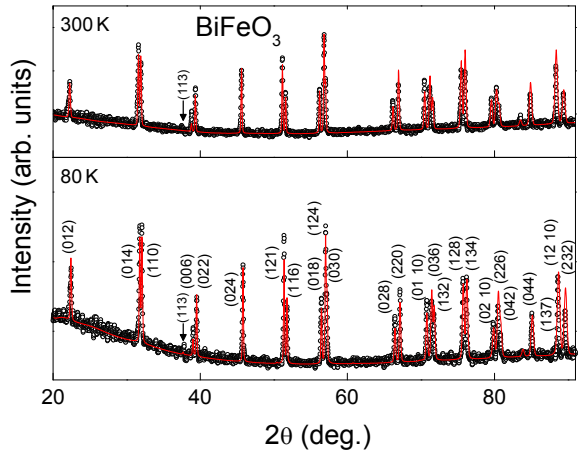

**Fig. 1.** Diffraction patterns of polycrystalline BiFeO$_3$ at room temperature (upper frame) and at 80 K (lower frame). In the lower frame the Bragg reflections are indexed in hexagonal setting. The solid lines are the results of Rietveld refinements. Please note that close to 37.5° the (113) reflection can be detected just above background, which is a superstructure reflection resulting from the doubling of the unit cell (indicated by arrows).

The samples have been structurally characterized with a STOE x-ray diffractometer utilizing Cu K$_\alpha$ radiation. The diffraction profile was collected between 10° < 2Θ < 90°. To search for possible structural phase transitions, measurements have been performed for temperatures between 80 K and 300 K. The diffraction patterns of crushed single crystals as well as those of polycrystalline samples show no anomalous temperature dependent shift or the splitting of Bragg peaks within experimental resolution that would indicate any type of structural phase transition in the temperature range investigated. Fig. 1 shows prototypical diffraction patterns as collected at 80 K and at 300 K, which at first sight look almost identical. Resolution-limited Bragg reflections were resolved throughout the angular range of the experiment. No traces of impurity phases could be detected above background. Fig. 1 documents that our samples reveal the proper R3c structure down to 80 K. The diffraction patterns also provide a hint of the weak (113) superstructure reflection (indicated by arrows in Fig. 1), which results from the doubling of the unit cell. This reflection is hardly above background but its traces can be observed in all diffractograms at all temperatures investigated in the course of this work.

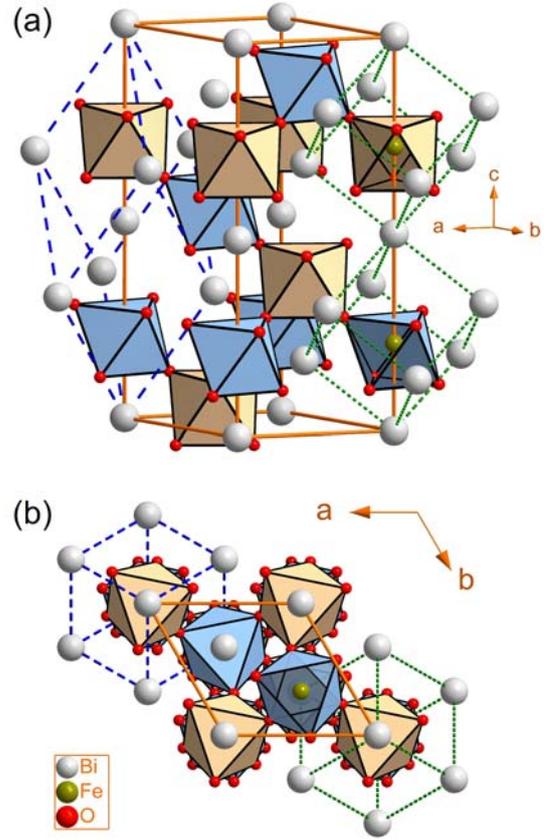

**Fig. 2.** Drawing of the structure of perovskite-type BiFeO$_3$. FeO$_6$ octahedra are indicated. Upper frame (a): Two octahedra (blue and yellow) constitute the rhombohedral unit cell indicated by the blue dashed lines. The hexagonal cell, containing 6 formula units, is indicated by the solid orange lines. The parent perovskite structure is shown by dotted green lines. Lower frame (b): Top view along the hexagonal c-axis. Here an iron ion is indicated within its (partly transparent) oxygen octahedron. The two oxygen octahedra within one rhombohedral unit cell are tilted by approximately 24° which implies that two subsequent octahedra are counter rotated by approximately 12° compared to the ideal perovskite structure.

For a more detailed analysis we performed a Rietveld refinement and found room-temperature lattice constants a = 0.5628(1) nm and α = 59.350(2)° in rhombohedral setting, corresponding to a = 0.55725 nm and c = 1.38529 nm when the hexagonal cell with six formula units is used instead. While the lattice constant a is slightly lower than that reported by Kubel and Schmid [43] the rhombohedral angles agree well within experimental uncertainties. These authors reported room temperature values of the lattice constants as determined in a monodomain single crystal as a = 0.56343 nm and α = 59.348°.

The temperature dependence of the lattice constant and of the rhombohedral angle α determined from our experiments (not shown), provide no evidence for a structural phase transition below the magnetic ordering temperature, in good agreement with the more intensive investigations presented in Ref. [11]. From the temperature dependence of the lattice constants, weak anomalies at the AFM ordering temperature have been reported by Bucci *et al*. [10] while the absence of magnetostrictive results has been deduced based on the experiments by Palewic *et al*. [11].

A pictorial view of the structure of BFO is given in Fig. 2. Here we also elucidate how rhombohedral setting with two formula units per unit cell and hexagonal setting with 6 formula units per unit cell are related. The figure clearly documents the strong buckling and tilting of the perovskite octahedral units which amount -12° and +12° in neighbouring octahedrons along the trigonal axis as compared to the ideal perovskite structure.

## 3 Experimental results and discussion

### 3.1 Magnetic properties

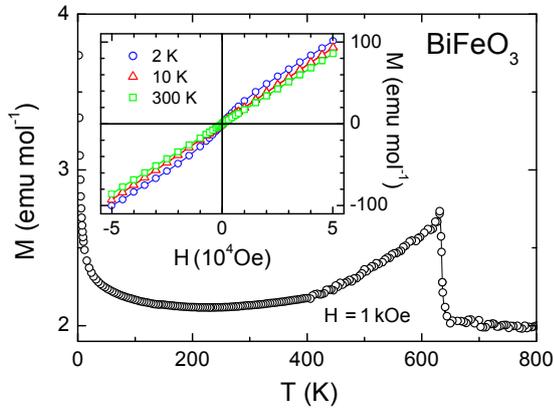

spins. No anomalies that might indicate further magnetic phase transitions can be detected in the temperature dependent susceptibility. The inset shows magnetisation measurements at 2, 10 and 300 K indicating the absence of a ferromagnetic hysteresis within experimental uncertainty. The weakly S-shaped magnetisation at 2 K results from the presence of a small but finite amount of defect spins related to the upturn in the susceptibility at low fields and at the lowest temperatures.

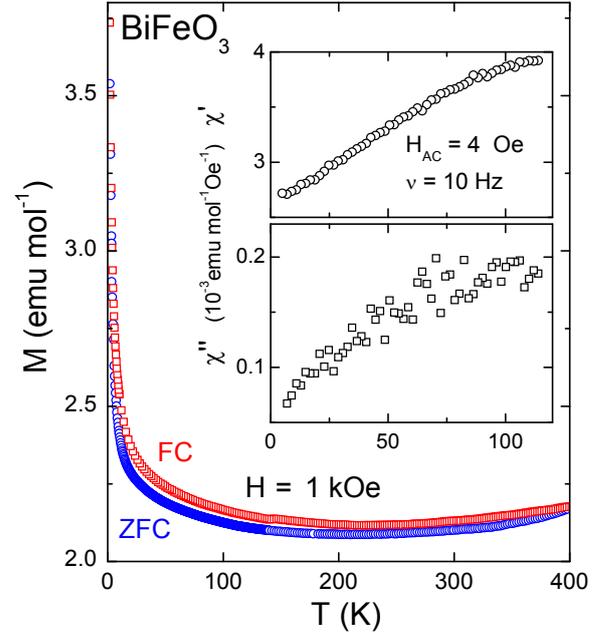

**Fig. 3.** Magnetisation *vs*. temperature of BiFeO$_3$ (large single crystal) as measured in an external field of 1 kOe for temperatures between 1.5 and 800 K. The midpoint of the step-like increase of the susceptibility at the onset of magnetic order is close to 635 K and is taken as estimate of the Néel temperature. The inset shows magnetisation *vs*. external fields taken at 2, 10 and 300 K. The slightly S-shaped curvature at the lowest temperatures close to zero fields comes from some residual magnetic impurities, which are also responsible for the upturn of the dc susceptibility towards 0 K. A similar figure was published in Ref. [51].

**Fig. 4.** Magnetisation of BiFeO$_3$ (large single crystal) as measured in external fields of 1 kOe under field cooling (FC) and zero-field cooling (ZFC) conditions. Within experimental uncertainties the two magnetisations coincide, providing no evidence for spin-glass like ordering. The insets show the results of ac susceptibility measurements taken with ac magnetic fields of 4 Oe and at a measuring frequency of 10 Hz. The temperature dependences of the real (upper frame) and imaginary part (lower frame) of the magnetic ac susceptibility are smooth and reveal no anomalies between 5 and 120 K.

All the magnetic properties that are discussed in the following were obtained on the large single crystal. Fig. 3 shows the magnetisation of BFO from 1.5 K to 800 K. A clear anomaly close to 635 K (midpoint of the step-like increase) signals the onset of long-range antiferromagnetic order. There is a striking and anomalous slow continuous decrease of the magnetisation, arriving at a room temperature susceptibility value of 2.2×10$^{-3}$ emu mol$^{-1}$ Oe$^{-1}$. The extremely low Curie-type increase towards 0 K definitely signals an almost impurity-free single crystal. Close to the magnetic ordering temperature our results compare nicely with those early reported by Roginskaya [6] and for temperatures below 150 K at least roughly compare with reports by Lebeugle *et al*. [16] on BFO single crystals. Our results reveal an even lower susceptibility and a weaker Curie-like upturn at low temperatures, indicative for a high-purity sample. If we analyze the low-temperature increase of the susceptibility as being due to defect Fe$^{3+}$ ions, we estimate approximately 0.7% defect

To search for possible spin-glass like ordering as has recently been reported by Singh *et al*. [19], we also performed field cooled (FC) and zero-field cooled (ZFC) susceptibility measurements. Representative results are depicted in Fig. 4. FC and ZFC cycles almost coincide excluding any type of spin glass ordering. The inset of Fig. 4 shows ac measurements at low fields (4 Oe) and low frequencies (10 Hz) in the temperature range where spin-glass freezing has been reported, which has been deduced from frequency dependent cusps in the ac susceptibility close to 30 K. Our results provide no experimental evidence for spin glass ordering. The real and imaginary parts of the dynamic susceptibility continuously decrease on decreasing temperature. This behaviour probably signals that the intrinsic susceptibility in BFO would decrease from the ordering temperature down to 0 K, but is in contrast to the dc measurements which are dominated by paramagnetic free impurity spins. This discrepancy between ac and dc magnetic susceptibility results, documented in Fig. 4, signals that at very low external magnetic fields a small ferromagnetic hysteresis should be present, which however is hard to resolve. Whether this weak ferromagnetism results from a marginal concentration of ferromagnetic clusters, whether it is of intrinsic origin driven by Dzyaloshinskii-Moriya type interactions as has been proposed by Ederer and Spaldin [44], or whether it results from a surface enhancement of the exchange interactions [45], currently cannot be answered. Irrespective of this fact, from these measurements we unambiguously conclude that no magnetic anomalies show up in BFO below room temperature and that reported phase or glass transitions most likely indicate impurity phases in samples of limited quality.

### 3.2 Thermodynamic measurements

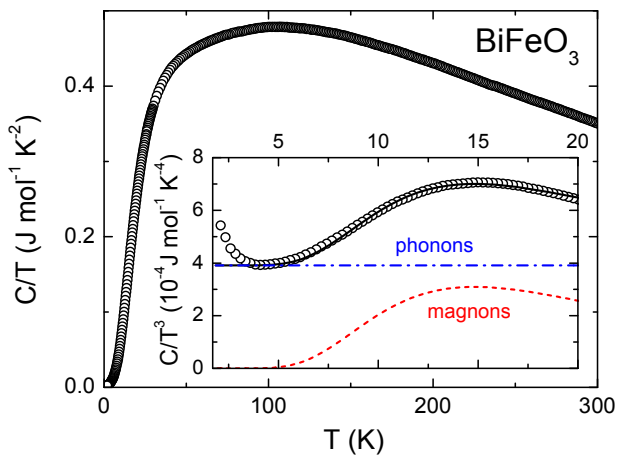

**Fig. 5.** Heat capacity of BiFeO$_3$ (large single crystal) for temperatures between 1.8 K and 300 K, plotted as C/T *vs.* T. The inset shows the low-temperature heat capacity C/T$^3$ *vs.* T on an enlarged scale. The experimental results have been fitted with a phonon-like T$^3$ and a gapped magnon contribution, described by an Einstein-like exponential term with constant gap energy (solid line). The phonon term (dash-dotted line) and the exponential Einstein term (dashed line) are also separately indicated. For temperatures below 20 K, both contributions are significant and can be unambiguously identified.

Fig. 5 shows the heat capacity of BiFeO$_3$ from 1.8 K to 300 K. For better representation of the experimental data, C/T *vs.* T is plotted. At first sight the heat capacity is dominated by phonon response and C/T reveals the characteristics of an insulating material. Again the absence of anomalies in the temperature dependence of the heat capacity signals the lack of any structural or magnetic phase transitions below room temperature. Even a spin glass transition would lead to a cusp-like contribution to the heat capacity just above the freezing temperature. However, at low temperatures the heat capacity provides the characteristic fingerprint of an ordered antiferromagnet. For a closer look, to specifically detect contributions of magnon modes, we made a more detailed analysis of the low-temperature data. One readily finds that a pure T$^3$ phonon term does not provide a satisfactory fit to the low-temperature specific heat. This means that neither a pure phonon model nor an additional T$^3$ term due to gapless magnons of an isotropic antiferromagnet describe the experimental data.

Obviously an exponential term, due to an anisotropy gap of the magnon excitations has to be considered in addition to the phonon contribution. The results are documented in the inset of Fig. 5 showing a plot of C/T$^3$. The best fit to the low temperature heat capacity is obtained using the sum of a T$^3$ phonon term, characteristic of the low-temperature Debye contribution in pure insulators, and of an exponential Einstein-like contribution due to gapped magnon modes (solid line in the inset). The two contributions are indicated also separately to document that both contributions have substantial weight and that at approximately 20 K the magnons and phonons equally contribute to the heat capacity of BFO. From this low-temperature fit we derive a Debye temperature of 292 K and a magnetic anisotropy gap of 73 K, corresponding to an energy of approximately 6 meV. Just recently the lowest magnon modes of BFO have been measured using Raman scattering techniques [46]. These authors found four magnon excitations located between 2.2 and 4.2 meV, significantly lower when compared to our result. But of course the specific-heat results correspond to a weighted average of magnon excitations over the complete Brillouin zone and over all frequencies. The complexity of a magnetic excitation spectrum in a transition-metal oxide, including crystal-field splitting, spin-orbit coupling and super-exchange interactions has just been documented by Kant *et al*. [47].

Towards the lowest temperatures investigated in the present experiments we found significant deviations of the specific heat from the canonical Debye-like behavior indicating a significantly lower temperature exponent of the heat capacity (inset of Fig. 5). In disordered or amorphous matter for T < 2 K the heat capacity follows approximately a linear temperature dependence. Deviations from the phonon-derived T$^3$ behavior are often observed even in single crystals and also may result from disordered regions in grain boundaries or domain walls.

### 3.3 Dielectric measurements

#### 3.3.1 Dielectric constants

In a first attempt we tried to measure the dielectric constants at audio and radio frequencies up to 1 MHz as function of temperature and frequency. Some representative results are shown in Fig. 6. These measurements were performed on the large single crystal. Here the upper frame shows the real part of the dielectric constant as function of temperature measured at frequencies between 1 Hz and 1 MHz. From first sight it is clear that the dielectric constants at ambient temperatures are strongly influenced by Maxwell-Wagner like relaxations as often observed in transition metal oxides [37]. This also means that the dielectric constants as measured at room temperature and low frequencies strongly depend on geometry and also on the type of contacts used in the experiments and thus allow no insight into the intrinsic dielectric properties. Very similar observations have been made by Kamba *et al*. [31]. In the large single crystal used in this experiment, at the highest frequencies the step-like increase of the dielectric constant due to Maxwell-Wagner effects is just shifted out of the temperature range investigated and thus the 1 MHz results provide a measure of the true and intrinsic dielectric constants of BFO. These high frequency data can be taken to search for anomalies due to structural or magnetic phase transitions coupled to the dielectric constant. In the inset of the upper frame of Fig. 6 we document an enlarged view of the temperature dependence of the dielectric constant as measured at 100 kHz and 1 MHz from 2 K up to room temperature. As expected in the absence of Maxwell-Wagner effects, there is no significant frequency dependence. The dielectric constant exhibits rather conventional temperature dependence, decreasing from approximately 120 at room temperature to 80 when approaching 0 K, typical for a canonical anharmonic solid. Measurements are shown on heating and cooling and some small anomalies may be suspected close to 170 and 240 K. We think that these anomalies are not significant and are well within the experimental uncertainties. For a final proof, we performed a number of similar experiments on small single crystals and were not able to unambiguously reproduce these anomalies.

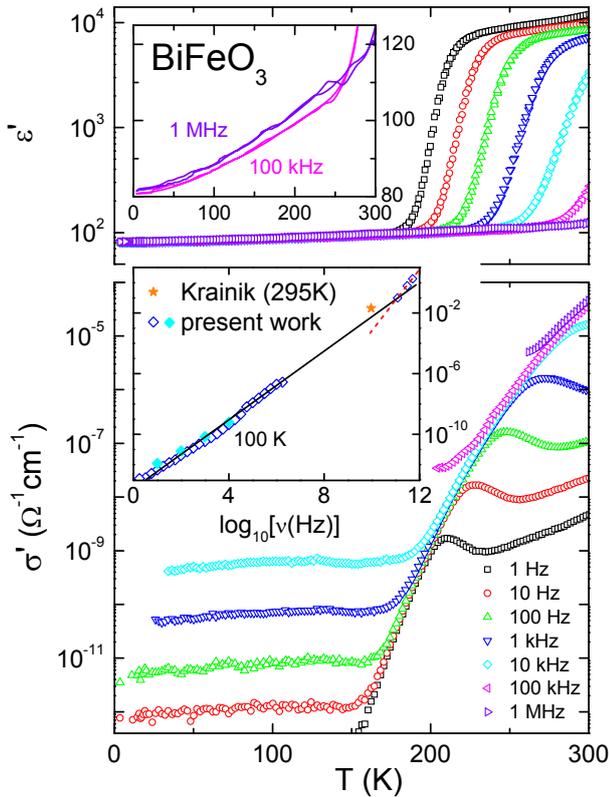

**Fig. 6.** Real part of the complex dielectric constant (upper frame) and real part of the conductivity (lower frame) *vs.* temperature as measured for frequencies between 1 Hz and 1 MHz on semi-logarithmic scales for temperatures between 1.5 K and 300 K. The frequencies are indicated in the figure. The inset in the upper frame shows the temperature dependence of the dielectric constant as measured at 100 kHz and 1 MHz on an enlarged linear scale. The lower inset shows the frequency dependence of the ac conductivity at 100 K (present work: full and empty diamonds). Also indicated is an early microwave result by Krainik [26] (star). The solid line indicates power-law behaviour of the conductivity with a frequency exponent of 1.15. The dashed line through the high-frequency results, obtained with a contact-free millimetre-wave technique, corresponds to a frequency exponent of 2.

It should be noted that in literature dielectric measurements at room temperature in BFO ceramics were reported, revealing the correct order of magnitude and obviously being only weakly influenced by Maxwell-Wagner effects (e.g., Ref. [21]). In these samples even at room temperature and above 10 kHz the dielectric constants are close to the intrinsic values signalling lower conductivities when compared to the results presented in Fig. 6. One could speculate that in some cases marginal doping of the ceramics decreases the conductivity substantially. It is well known that decreasing conductivity shifts the step of the Maxwell-Wagner relaxation towards higher temperatures, allowing to measure intrinsic values of the dielectric constants at room temperature [48]. In these experiments, small anomalies close to 50 K, 140 K and around 200 K have been reported. The low-temperature anomalies have been traced back to a spin glass transition (50 K) and a spin-reorientation transition (140 K). The transition close to 200 K seems to be rather elastic in nature and only weakly coupled to the lattice. On the basis of our results, including magnetic susceptibility and heat capacity, we raise some doubt on the occurrence of these phase transitions in high-purity samples. However, we certainly cannot definitely exclude their existence in case these transitions leave minor fingerprints in the magnetic, dielectric, or thermodynamic experiments.

From our measurements we deduce a low-temperature intrinsic value of $\varepsilon \sim 80$, which however is not very exact: In similar experiments, Kamba *et al.* [31] deduced a value close to 40 while in Ref. [21], depending on sample, values between about 40 and 90 were reported. We performed a number of additional low-temperature dielectric measurements on small single crystals and found an average value of $\varepsilon \sim 50$. Overall these measurements suffer from large scatter mainly due to ill defined geometries, stray capacitance and ac conductivity contributions. We feel that for obtaining absolute values of the dielectric constants, one is well advised to rely on contact free measurements at high frequencies as discussed later.

The lower frame of Fig. 6 shows the conductivity *vs.* temperature using semi-logarithmic scales. The temperature-independent conductivities towards low temperatures mainly correspond to ac conductivity contributions due to hopping charge carriers. This ac conductivity roughly shows a $\nu^1$ dependence and the low-temperature frequency dependence of the conductivity is documented in the lower inset of Fig. 6. The full diamonds correspond to the measurements documented in the lower frame of Fig. 6. The empty diamonds are measurements from a small single crystal. We see that both ac conductivities are of the same order of magnitude. The solid line represents an increase of the ac conductivity with a frequency exponent of approximately 1.15. A similar super-linear power law, whose origin remains to be explained, is quite universally observed in many different materials, including also transition-metal oxides [49].

Towards higher temperatures the conductivities $\sigma'(T)$ in Fig. 6 merge into a frequency-independent curve, which represents the intrinsic dc conductivity, before towards room temperatures Maxwell-Wagner contributions start to dominate, leading to strong deviations from the intrinsic behaviour. By performing measurements above room temperature, we tried to follow the intrinsic dc conductivity towards higher temperatures. The results are documented in Fig. 7. This figure suggests that we were able to measure the true dc conductivity $\sigma_{dc}$ of BFO from 130 K at least up to 600 K. The values increase from $10^{-13}$ $\Omega^{-1}$cm$^{-1}$ at 120 K to $10^{-1}$ $\Omega^{-1}$cm$^{-1}$ at 600 K close to the antiferromagnetic phase transition. At higher temperatures, $\sigma'(T)$ levels off at a nearly constant value, which seems unlikely to represent the intrinsic behaviour of BFO. In Fig. 7, we also included results for the dc conductivity (stars) deduced by fitting the frequency dependence of $\sigma'$ at different temperatures using an equivalent-circuit model [37]. Please note that only in a rather limited temperature range measurements at a given frequency provide the intrinsic values of the dc conductivity. $\sigma_{dc}(T)$ can be well described by an Arrhenius fit, indicated as solid line, yielding an energy barrier E = 390 meV. For clarity we plotted the same data in an Arrhenius representation in the inset of Fig. 7, revealing straight-line behaviour of the intrinsic dc conductivity. The results of Fig. 7 are not in agreement with those reported, e.g., by Selbach *et al.* [50] who found a value of the conductivity close to $10^{-5}$ $\Omega^{-1}$cm$^{-1}$ at the magnetic ordering temperature, substantially lower than our value. However, it is clear that BFO samples have different amounts of impurity phases and defect ions or reveal different oxygen stoichiometries. Accordingly they will reveal different conductivities depending on the different doping levels and it seems most likely that even up to high temperatures BFO is not dominated by the intrinsic conductivity that would result from the optical gap, but merely by unavoidable impurity contributions and defect centres.

experimental uncertainties. Interestingly, neither the dielectric constant nor the loss tangent show temperature dependences in the FE state that are characteristic for the dielectric properties of ferroelectrics. Neglecting the number of small step-like increases of the dielectric constant on increasing temperature observed in that work between room temperature and 1100 K, the onset of magnetic order can hardly be detected, raising doubt on strong magnetoelectric coupling. This is specifically true for the temperature dependence of the dielectric loss, which is temperature independent revealing no anomalies within experimental uncertainty.

Using millimetre wave spectroscopy, we performed measurements of the transmission and phase shift in the GHz range for temperatures from 70 K to 300 K. The results are shown in Fig. 8. At 4 cm$^{-1}$ (~ 120 GHz), we found a dielectric constant with a value close to 53 at room temperature, which decreases with decreasing temperatures. These values are close to those reported by Krainik [26] and are also close to values as determined by FIR spectroscopy [51]. Hence we believe that this value is close to the intrinsic dielectric constant. These measurements also provide no clear experimental evidence for anomalies as reported recently [21]. However, we note that on the basis of this data and on the data presented in the inset of Fig. 6, it is hard to exclude minor dielectric anomalies.

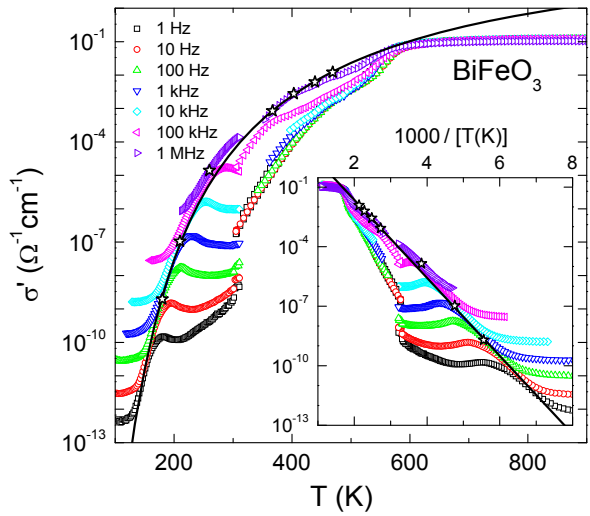

**Fig. 7.** Real part of the ac conductivity for temperatures between 100 and 900 K as measured for frequencies between 1 Hz and 1 MHz. The open stars represent results of the dc conductivity as determined from fits of $\sigma'(\nu)$ using an equivalent circuit description [37]. In the inset, the same experimental data are plotted in an Arrhenius-type representation. A fit utilizing a pure Arrhenius behaviour is indicated as solid line in both frames.

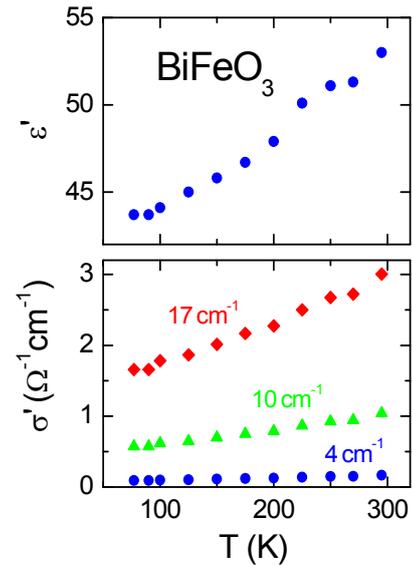

**Fig. 8.** Dielectric constant (upper frame) and conductivity (lower frame) *vs.* temperature in BiFeO$_3$ as measured by contact free millimetre-wave techniques on the large single crystal at frequencies as indicated in the figure. The upper frame was taken from Ref. [51].

Figs. 7 and 8 also document that the intrinsic dielectric properties can best be derived from contact-free measurements. Early microwave experiments have been performed by Krainik *et al.* [26]. These authors measured the real and imaginary parts of the dielectric constant at 9.4 GHz from room temperature up to 1150 K. They found a linear increase of the dielectric constant from approximately 45 at room temperature to approximately 150 just below the FE phase transition. The loss tangent was close to 0.09 at all temperatures. At 1120 K a small peak in the temperature-dependent dielectric constant and a drop in thermal expansion have been taken as evidence for the FE transition. These authors speculate on the observation of a large number of subsequent phase transitions over the entire temperature range. However, all observed "anomalies" seem to be within

We independently also determined the ac conductivity (lower frame of Fig. 8), as measured at different frequencies. It also decreases on decreasing temperature and strongly depends on frequency. The frequency dependence of the conductivity at 100 K is plotted in the lower inset of Fig. 6 in a double logarithmic representation (three uppermost diamonds) and compared to the results obtained at low frequencies, discussed above. In this inset the contact-free result as obtained by Krainik *et al.* [26] at room temperature is also included. Obviously, the low-frequency superlinear extrapolation nicely meets the high-frequency results. However, our submillimetre results definitely indicate stronger frequency dependence. It might well be that the high-frequency conductivity points are already in a regime where photon-assisted tunnelling has to be considered, which leads to a frequency exponent of 2 [52,53].

### 3.3.2 Ferroelectric polarization

First measurements of FE hystereses in BFO have been reported by Teague *et al.* [24]. They found a spontaneous polarization of 3.5 $\mu C/cm^2$ in fields of 55 kV/cm along the pseudocubic [100] direction at liquid helium temperature. At room temperature the samples were too conductive to allow for hysteresis experiments. Later on, Wang *et al.* [4] found FE hystereses in thin films with a polarization of 55 $\mu C/cm^2$ in fields of 500 kV/cm. Finally, Lebeugle *et al.* [16] determined FE hystereses in a 40 µm thick single crystal along the hexagonal [012] axis (corresponding to the pseudocubic [100] direction) in fields of 125 kV/cm. The saturation polarization was 35 $\mu C/cm^2$, the coercitivity 15 kV/cm. The observation of FE hystereses proved that the single crystals used in that work have a high resistivity, even at room temperature. The authors gave an estimate of $6 \times 10^{10}$ Ωcm, which is five decades higher than the resistivity of the samples investigated in the present work (see Fig. 6).

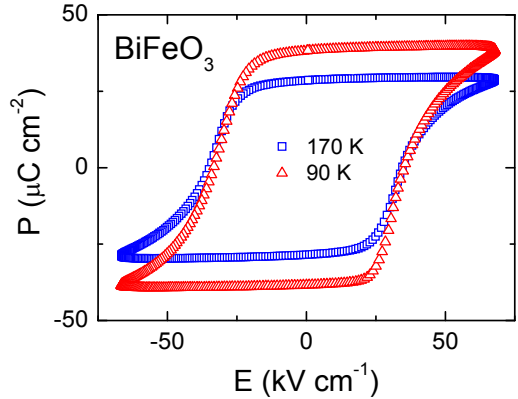

**Fig. 9.** Ferroelectric polarization *vs.* external electric field as measured at 90 K and 170 K at a measuring frequency of 6 Hz.

In the present work we performed polarization experiments on a large number of thin single crystals. From Fig. 6 it is clear that we would not be able to observe reliable hysteresis loops at room temperature at conventional measuring frequencies, as there Maxwell-Wagner effects dominate and the conductivity is too high. Instead we had to perform these experiments below 200 K, where intrinsic values of the dielectric constant are observed. In addition, we found that FE polarization in BFO cannot be observed in external electric fields lower than 20 kV/cm, making the availability of thin samples a prerequisite for obtaining reliable results. In Fig. 9 we show two typical hysteresis loops, measured below 200 K in external fields up to 60 kV/cm and at frequencies of 6 Hz. These hystereses measurements have been performed with an excitation voltage of the applied electric field up to 250 V on a crystal with approximately 0.2 $mm^2$ and a thickness of 40 µm. The remanent polarization which is almost equal to the saturation polarization is of the order of 40 $\mu C/cm^2$ and the coercitivity of the order of 35 kV/cm. The FE hysteresis loops exhibit weak temperature dependence, but overall remain almost constant for temperatures below 200 K. In a large number of slightly thicker samples we found hystereses with remanent polarizations of the order of 5 - 10 $\mu C/cm^2$ only. If these low values result from the orientation dependence of the polarization or from the fact that we were not able to saturate the polarization in thicker (> 50 µm) bulk crystals is unclear.

Concerning the results presented in Fig. 9, if we assume that the platelet-shaped single crystals have the largest faces corresponding to the (010) pseudo cubic planes of the parent perovskite structure [16], respectively the (012) plane of the hexagonal unit cell (see Fig. 2), our results compare excellently with those of Lebeugle *et al.* [16]. They would correspond to a full saturation polarization of 70 $\mu C/cm^2$ along the hexagonal [001] direction, in good agreement with the measurements on single crystals [16] and epitaxial thin films [4] and also with first principles calculations [54].

## 4 Summary and conclusions

During the course of this work we have synthesized high-purity BFO ceramics as well as high-purity single crystals. We also were supplied with a large and high-quality single crystal by V. A. Murashov. The most important results can be summarized as follows:

i) The magnetic phase transition is close to 635 K, characterized by a step like increase of the magnetic susceptibility and a slow continuous decrease towards low temperatures.

ii) Detailed dc FC and ZFC experiments as well as ac measurements of the magnetic susceptibility, performed at low fields, provide no evidence for any further magnetic phase transition, spin reorientation or spin glass transition at lower temperatures. Obviously, at 635 K FE BFO undergoes a transition into a stable spiral spin phase. Spin reorientation transitions are sometimes observed in ferrimagnets with two magnetic sublattices, mainly driven by rare earth exchange interactions. It is unclear how spin reorientation could be explained in BFO.

iii) Even in these high purity samples we found remarkable differences in the temperature dependences of ac and dc susceptibility measurements. These differences probably result from a marginal concentration of ferromagnetic clusters in the sample. However, we can not rule out (very) weak intrinsic ferromagnetism.

iv) Below room temperature the heat capacity is dominated by the normal phonon response. Again no indications of structural or magnetic phase transitions were detected. At low temperatures (T < 30 K) we detect significant contributions from magnon excitations with an anisotropy gap of approximately 6 meV.

v) The low-frequency (< 1 MHz) dielectric response at elevated temperatures (T > 300 K) is dominated by Maxwell-Wagner like relaxations. But no significant dielectric anomalies can be detected below room temperature following the temperature dependence of the intrinsic dielectric constants measured at MHz frequencies. However, we have to admit that small anomalies cannot be fully excluded. Neglecting Maxwell-Wagner effects, the dielectric response is absolutely dispersion free at least up to room temperature. The conductivity reveals an exponential increase between 130 K and 600 K and can be described by a single activation energy of 390 meV.

vi) In addition, we performed contact-free measurements of the dielectric constant at 120 GHz between room temperature and 80 K. From these experiments we deduce a room temperature value of the dielectric constant of 53 and a smooth decrease down to 42 close to 80 K. We provide arguments that 53 come close to the intrinsic room temperature dielectric constant of BFO.

vii) The conductivity at low temperatures exhibits a clear super-linear power-law with a frequency exponent of 1.15. In the GHz range a transition to stronger frequency dependence with an exponent close to 2 is observed.

viii) In ultra-thin single crystals and for temperatures below 200 K we observed a FE polarization of the order of 40 µC/cm$^2$ in electric fields of 50 kV/cm with a coercitivity of approximately 35 kV/cm.

This work was supported by the Deutsche Forschungsgemeinschaft via the collaborative research centre SFB 484 (Augsburg) and the transregional collaborative research center TRR80 (Augsburg, Munich). J. Lu acknowledges support from the China Scholarship council for his oversea PhD studies. We thank V. A. Murashov for providing large and high quality BFO single crystals.

**References**

1. G. A. Smolenskii, I. E. Chupis, Sov. Phys. Usp. **25**, 475 (1982)
2. M. Fiebig, J. Phys. D **38**, R123 (2005)
3. S. W. Cheong, M. Mostovoy, Nature Mater. **6**, 13 (2007); R. Ramesh, N. A. Spaldin, Nature Mater. **6**, 21 (2007)
4. J. Wang, J. B. Neaton, H. Zheng, V. Nagarajan, S. B. Ogale, B. Liu, D. Viehland, V. Vaithyanathan, D. G. Schlom, U. V. Waghmare, N. A. Spaldin, K. M. Rabe, M. Wuttig, R. Ramesh, Science **299**, 1719 (2003)
5. R. T. Smith, G. D. Achenbach, R. Gerson, W. J. James, J. Appl. Phys. **39**, 70 (1968)
6. Yu. E. Roginskaya, Y. Y. Tomashpolskii, Y. N. Venevtsev, V. M. Petrov, G. S. Zhdanov, Sov. Phys. JETP **23**, 47 (1966)
7. S. V. Kiselev, G. S. Zhdanov, R. P. Ozeyov, Sov. Phys. Dokl. **7**, 742 (1963)
8. P. Fischer, M Polomska, I Sosnowska, M Szymanski, J. Phys. C: Solid State Phys. **13**, 1931 (1980)
9. J. M. Moreau, C. Michel, R. Gerson, W. J. James, J. Phys. Chem. Sol. **32**, 1315 (1971)
10. J. D. Bucci, B. K. Robertson, W. J. James, J. Appl. Cryst. **5**, 187 (1972)
11. A. Palewicz, R. Przeniosło, I. Sosnowska, A. W. Hewat, Acta Cryst. B **63**, 537 (2007); A. Palewicz, I. Sosnowska, R. Przeniosło, A. W. Hewat, Acta Phys. Polonica A **117**, 296 (2010)
12. A. J. Jacobson, B. E. F. Fender, J. Phys. C: Solid State Phys. **8**, 844 (1975)
13. I. Sosnowska, T. Peterlin-Neumaier, E. Steichele, J. Phys. C: Solid State Phys. **15**, 4835 (1982)
14. R. Przeniosło, A. Palewicz, M. Regulski, I. Sosnowska, R. M. Ibberson, K. S. Knight, J. Phys. Condens. Matter **18**, 2069 (2006)
15. Yu. F. Popov, A. K. Zvezdin, G. P. Vorob´ev, A. M. Kadomtseva, V. A. Murashev, D. N. Rakov, JETP Lett. **57**, 69 (1993); A. M. Kadomtseva, A. K. Zvezdin, Yu. F. Popov, A. P. Pyatakov, G. P. Vorob´ev, JETP Lett. **79**, 571 (2004).
16. D. Lebeugle, D. Colson, A. Forget, M. Viret, P. Bonville, J. F. Marucco, S. Fusil, Phys. Rev. B **76**, 024116 (2007)
17. F. Söffge, W. v. Hörsten, J. Magn. Magn. Mater. **59**, 135 (1986)
18. S. Nakamura, S. Soeya, N. Ikeda, M. Tanaka, J. Appl. Phys. **74**, 5652 (1993)
19. M. K. Singh, W. Prellier, M. P. Singh, R. S. Katiyar, J. F. Scott, Phys. Rev. B **77**, 144403 (2008)
20. M. K. Singh, R. S. Katiyar, J. F. Scott, J. Phys.: Condens. Matter **20**, 252203 (2008)
21. S. A. T. Redfern, C. Wang, J. W. Hong, G. Catalan, J. F. Scott, J. Phys.: Condens. Matter **20**, 452205 (2008)
22. M. Cazayous, Y. Gallais, A. Sacuto, R. de Sousa, D. Lebeugle, D. Colson, Phys. Rev. Lett. **101**, 037601 (2008)
23. G. Catalan, J. F. Scott, Adv. Mater. **21**, 2463 (2009)
24. J. R. Teague, R. Gerson, W. J. James, Solid State Commun. **8**, 1073 (1970)
25. H. Naganuma, S. Okamura, J. Appl. Phys. **101**, 09M103 (2007); H. Naganuma, T. Okubo, S. Sekiguchi, Y. Ando, S. Okamura, J. Appl. Phys. **105**, 07D903 (2009)
26. N. N. Krainik , N. P. Khuchua, V. V. Zhdanova, V. A. Evseev, Sov. Phys. Solid State **8**, 654 (1966)
27. M. Polomska, W. Kaczmarek, Z. Pajak, Phys. Stat. Sol. A **23**, 567 (1974)
28. M. M. Kumar, V. R. Palkar, K. Srinivas, S. V. Suryanarayana, Appl. Phys. Lett. **76**, 2764 (2000)
29. Y.-K. Jun, W.-T. Moon, C.-M. Chang, H.-S. Kim, H. S. Ryu, J. W. Kim, K. H. Kim, S.-H. Hong, Solid State Commun. **135**, 133 (2005)
30. R. Mazumder, S. Ghosh, P. Mondal, D. Bhattacharya, S. Dasgupta, N. Das, A. Sen, A. K. Tyagi, M. Sivakumar, T. Takami, H. Ikuta, J. Appl. Phys. **100**, 033908 (2006)
31. S. Kamba, D. Nuzhnyy, M. Savinov, J. Sebek, J. Petzelt, J. Prokleska, R. Haumont, J. Kreisel, Phys. Rev. B **75**, 024403 (2007)
32. J.-C. Chen, J-M. Wu, Appl. Phys. Lett. **91**, 182903 (2007)
33. S. Hunpratub, P. Thongbai, T. Yamwong, R. Yimnirun, S. Maensiri, Appl. Phys. Lett. **94**, 062904 (2009)
34. V. R. Palkar, J. John, R. Pinto, Appl. Phys. Lett. **80**, 1628 (2002)
35. S.-H. Lim, M. Murakami, J. H. Yang, S.-Y. Young, J. Hattrick-Simpers, M. Wuttig, L. G. Salamanca-Riba, I. Takeuchi, Appl. Phys. Lett. **92**, 012918 (2008)
36. X.-Y. Zhang, Q. Song, F. Xu, C. K. Ong, Appl. Phys. Lett. **94**, 022907 (2009)
37. P. Lunkenheimer, V. Bobnar, A. V. Pronin, A. I. Ritus, A. A. Volkov, A. Loidl, Phys. Rev. B **66**, 052105 (2002)
38. Z. V. Gabbasov, M. D. Kuz'min, A. K. Zvezdin, I. S. Dubenko, V. A. Murashov, D. N. Rakov, I. B. Krynetsky, Phys. Lett. A **158,** 491 (1991)
39. U. Schneider, P. Lunkenheimer, A. Pimenov, R. Brand, A. Loidl, Ferroelectrics **249**, 89 (2001)
40. J. F. Scott, J. Phys.: Condens. Matter **20**, 021001 (2008)
41. A. Loidl, S. Krohns, J. Hemberger, P. Lunkenheimer, J. Phys.: Condens. Matter **20**, 191001 (2008)

42. B. Gorshunov, A. A. Volkov, I. Spektor, A. M. Prokhorov, A. Mukhin, M. Dressel, S. Uchida, A. Loidl, Int. J. Infrared and Millimeter Waves **26**, 1217 (2005)
43. F. Kubel, H. Schmid, Acta Cryst. B **46**, 698 (1990)
44. C. Ederer, N. A. Spaldin, Phys. Rev. B **71**, 060401(R) (2005)
45. J. M. Wesselinowa, I. Apostolova, J. Appl. Phys. **104**, 084108 (2008).
46. P. Rovillain *et al*., arXiv 0905.1891
47. Ch. Kant, T. Rudolf, F. Schrettle, F. Mayr, J. Deisenhofer, P. Lunkenheimer, M. V. Eremin, A. Loidl, Phys. Rev. B **78**, 245103 (2008)
48. S. Krohns, J. Lu, P. Lunkenheimer, V. Brizé, C. Autret-Lambert, M. Gervais, F. Gervais, F. Bourée, F. Porcher, A. Loidl, Eur. Phys. J. B (2009), in press
49. P. Lunkenheimer, A. Loidl, Phys. Rev. Lett. **91**, 207601 (2003)
50. S. M. Selbach, T. Tybell, M.-A. Einarsrud, T. Grande, Adv. Mater. **20**, 3692 (2008)
51. Jun Lu, M. Schmidt, P. Lunkenheimer, A. Pimenov, A. A. Mukhin, V. D. Travkin, A Loidl, J. Phys.: Conf. Series **200**, 012106 (2010).
52. N. F. Mott, Philos. Mag. **22**, 7 (1970)
53. E. Helgren, N. P. Armitage, G. Grüner, Phys. Rev. Lett. **89**, 246601 (2002)
54. J. B. Neaton, C. Ederer, U. V. Waghmare, N. A. Spaldin, K. M. Rabe, Phys. Rev. B **71**, 014113 (2005); P. Ravindran, R. Vidya, A. Kjekshus, H. Fjellvag, O. Eriksson, Phys. Rev. B **74**, 224412 (2006); H.-J. Feng, F.-M. Liu, Phys. Lett. A **372**, 1904 (2008)